\pdfoutput=1

\documentclass[9pt,twocolumn,twoside]{newclass}

\title{Deep-STORM: super-resolution single-molecule microscopy by deep learning}

\date{}

\author[1,2,*]{Elias Nehme}
\author[2]{Lucien E. Weiss}
\author[1]{Tomer Michaeli}
\author[2]{Yoav Shechtman}
\affil[1]{Department of Electrical Engineering, Technion, 32000 Haifa, Israel}
\affil[2]{Department of Biomedical Engineering, Technion, 32000 Haifa, Israel}
\affil[*]{Corresponding author: seliasne@campus.technion.ac.il}

\begin{document}

\maketitle

\section*{Abstract}
We present an ultra-fast, precise, parameter-free method, which we term Deep-STORM, for obtaining super-resolution images from stochastically-blinking emitters, such as
fluorescent molecules used for localization
microscopy. Deep-STORM uses a deep convolutional neural network that can be trained on simulated data or experimental measurements, both of which are demonstrated. The method achieves state-of-the-art resolution under challenging signal-to-noise conditions and high emitter densities, and is significantly faster than existing approaches. Additionally, no prior information on the shape of the underlying structure is required, making the method applicable to any blinking data-set. We validate our approach by super-resolution image reconstruction of simulated and experimentally obtained data.

\section{Introduction}
In conventional microscopy, the spatial resolution of an image is bounded by Abbe’s diffraction limit, corresponding to approximately half the optical wavelength. Super resolution methods, e.g. stimulated emission depletion (STED) \citep{Hell:94,Klar:99}, structured illumination micrscopy (SIM) \citep{Neil:97,doi:10.1080/713817795,Gustafsson2000}, and localization microscopy, namely photo-activated localization microscopy ((F)PALM) \citep{Betzig,HESS20064258} and stochastic optical reconstruction microscopy (STORM) \citep{Rust2006} have revolutionized biological imaging in the last decade, enabling the observation of cellular structures at the nanoscale \citep{SAHL2013778}. Localization microscopy relies on acquiring a sequence of diffraction-limited images, each containing point-spread functions (PSFs) produced by a sparse set of emitting fluorophores. Next, the emitters are localized with high precision. By combining all of the recovered emitter positions from each frame, a super-resolved image is produced with resolution typically an order of magnitude better than the diffraction limit (down to tens of nanometers).

In localization microscopy, regions with a high density of overlapping emitters pose an algorithmic challenge. This emitter-sparsity constraint leads to a long acquisition time (seconds to minutes), which limits the ability to capture fast dynamics of sub-wavelength processes within live cells. Various algorithms have been developed to handle overlapping PSFs. Existing classes of algorithms are based on sequential fitting of emitters, followed by subtraction of the model PSF \citep{HoGBOM1974,Serge2008,Qu2004,Gordon}; blinking statistics \citep{Dertinger2009,Cox2012,Gustafsson2016}; sparsity \citep{Holden2011,Zhu2012,Barsic2014,Min2015,Gazagnes2017,Hugelier2016,solomon2017sparcom}; multi-emitter maximum likelihood estimation \citep{Huang2011}; or even single-image super-resolution by dictionary learning \citep{Mutzafi2015,Mutzafi2017}. While successful localization of densely-spaced emitters has been demonstrated, all existing methods suffer from two fundamental drawbacks: data-processing time and sample-dependent paramter tuning. Even accelerated sparse-recovery methods such as CEL0 \citep{Gazagnes2017}, which employs the fast FISTA algorithm \citep{Beck2009}, still involve a time-consuming iterative procedure, and scale poorly with the recovered grid size. In addition, current methods rely on parameters that balance different tradeoffs in the recovery process. These need to be tuned carefully through trial and error to obtain satisfactory results; ergo, requiring user expertise and tweaking-time.

Here we demonstrate precise, fast, parameter-free, super-resolution image reconstruction by harnessing Deep-Learning. Convolutional neural networks have shown impressive results in a variety of image processing and computer-vision tasks, such as single-image resolution enhancement \citep{Dong2016,Kim2016,Wang2016,NIPS2016_6172,Rivenson2017} and segmentation \citep{Long2015,ronneberger2015u,Noh2015}. In this work, we employ a fully convolutional neural network for super-resolution image reconstruction from dense fields of overlapping emitters. Our method, dubbed Deep-STORM, does not explicitly localize emitters. Instead, it creates a super-resolved image from the raw data directly. The net produces images with reconstruction resolution comparable or better than existing methods; furthermore, the method is extremely fast, and our software can leverage GPU computation for further enhanced speed. Moreover, Deep-STORM is parameter free, requiring no expertise from the user, and is easily implemented for any single-molecule dataset. Importantly, Deep-STORM is general and does not rely on any prior knowledge of the structure in the sample, unlike recently demonstrated, single-shot image enhancement by Deep-Learning \citep{Weigert236463}.

\section{Methods}
\subsection{Deep Learning}
In short, Deep-STORM utilizes an artificial neural net that receives a set of frames of (possibly very dense) point emitters and outputs a set of super-resolved images (one per frame), based on prior training performed on simulated or experimentally obtained images with known emitter positions. The output images are then summed to produce a single super-resolved image.

\begin{figure}[htbp]
\centering
\includegraphics[width=\linewidth]{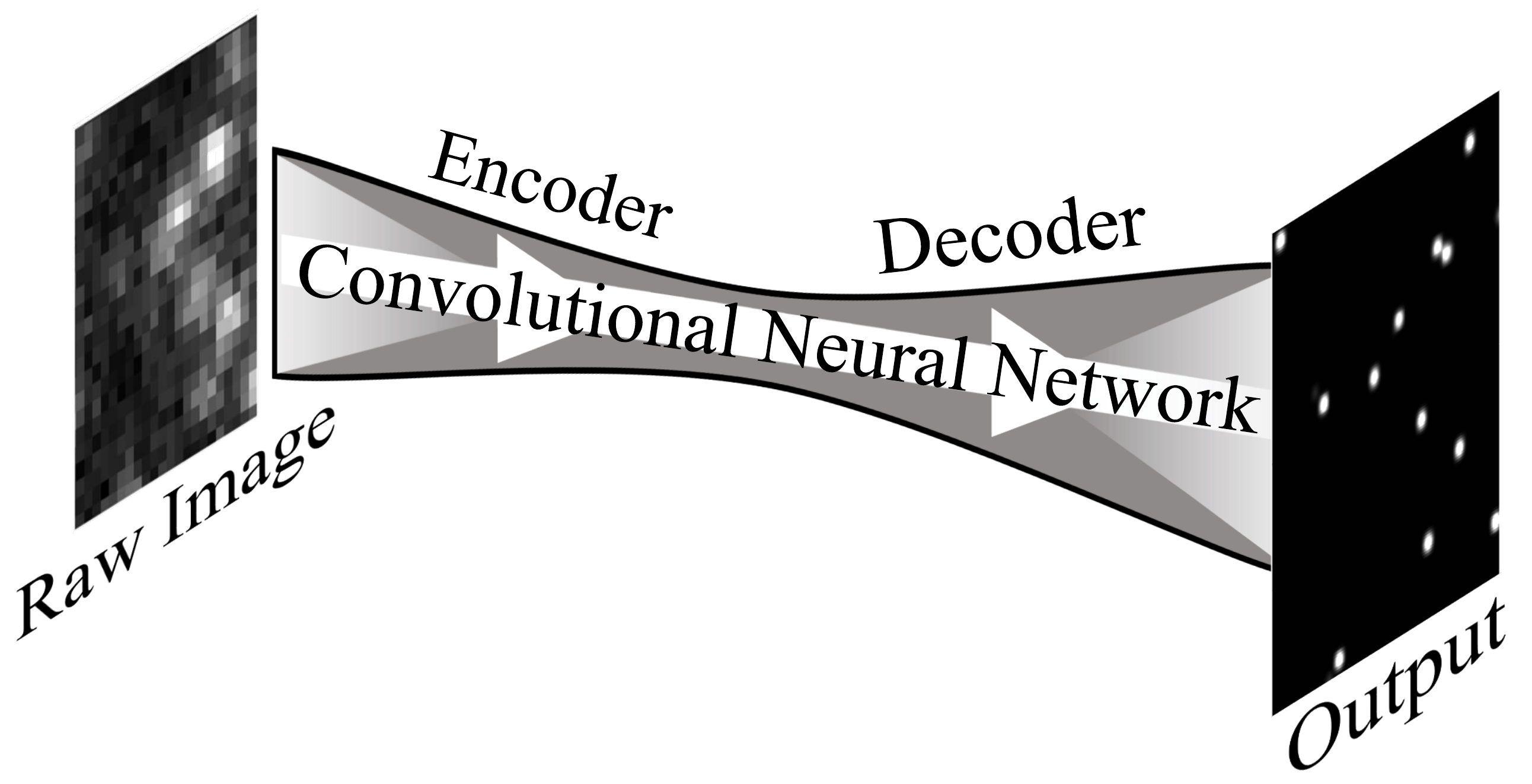}
\caption{Network architecture. A set of low-resolution diffraction-limited images of stochastically blinking emitters is fed into the network to produce reconstructed high-resolution images. The resulting outputs are then summed to generate the final super-resolved image.}
\label{fig:Architecture}
\end{figure}

\subsubsection{Architecture}
The net-architecture is based on a fully convolutional encoder-decoder network and was inspired by previous work on cell counting \citep{Xie2016}. The  network (Figure \ref{fig:Architecture}) first encodes the input intensity-image into a dense, aggregated feature-representation, through three $3\times3$ convolutional layers with increasing depth, interleaved with $2\times2$ max-pooling layers (Supplementary Information). The result is an encoded representation of the data. Afterwards, in the decoding stage, the spatial dimensions are restored to the size of the input image through three successive deconvolution layers, each consisting of $2\times2$ upsampling, interleaved with $3\times3$ convolutional layers with decreasing depth. Convolutional layers, for both encoding and decoding, refer to a composite of convolution filters, followed by Batch Normalization (BN) \citep{pmlr-v37-ioffe15}, and then a ReLU non-linearity \citep{Maas2013}. Since only $3\times3$ filters are used, the resulting architecture size is relatively small, only $1.3M$ trainable parameters. The final pixel-wise prediction (super-resolution frame) is created using a depth-reducing $1\times1$ convolutional filter with a linear-activation function. See supplementary information for architecture details. 

\subsubsection{Training}
\begin{figure}[htbp]
\centering
\includegraphics[width=\linewidth]{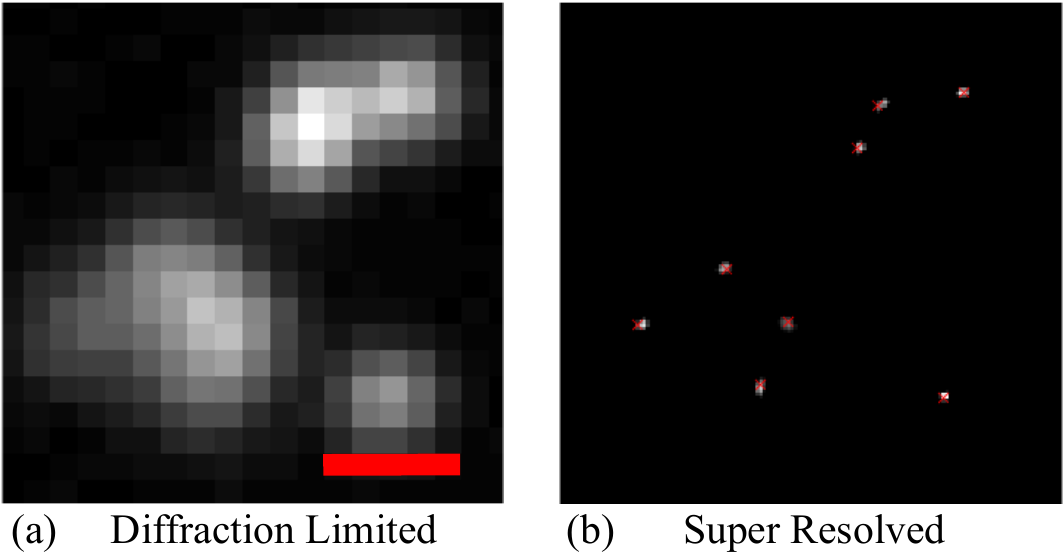}
\caption{Simulated dense emitters. (a) Low-resolution image. Scale bar is $0.5 \ \mu m$. (b) Deep-STORM prediction on a $12.5 \ nm$ grid with ground truth emitter locations overlaid as cross marks on top.}
\label{fig:DenseEmitters}
\end{figure}

Given the camera specifications, PSF model, approximate signal-to-noise ratio (SNR), and the expected emitter density, twenty $64\times64$ pixel images containing randomly positioned emitters are simulated using the ImageJ \citep{Rueden2017,Schindelin2012} ThunderSTORM plugin \citep{Ovesny2014}. From each frame we extract $500$ random $26\times26$ regions and their respective ground truth $xy$ emitter-positions. To generate the final training examples we upsample each region by a factor of 8, and project the appropriate emitter positions on the high-resolution grid. The result is a set of $10K$  pairs of upsampled low-resolution regions ($208\times208$ pixels) alongside images with spikes at the ground truth positions, used as training examples. Each region is normalized using the mean and averaged standard deviation (per-region) of the dataset without additional data augmentation. An example training input-image and the corresponding output (after training) are shown in Figure \ref{fig:DenseEmitters}.

\begin{figure*}[htbp]
\centering
\includegraphics[width=0.72\linewidth, height=0.55\linewidth]{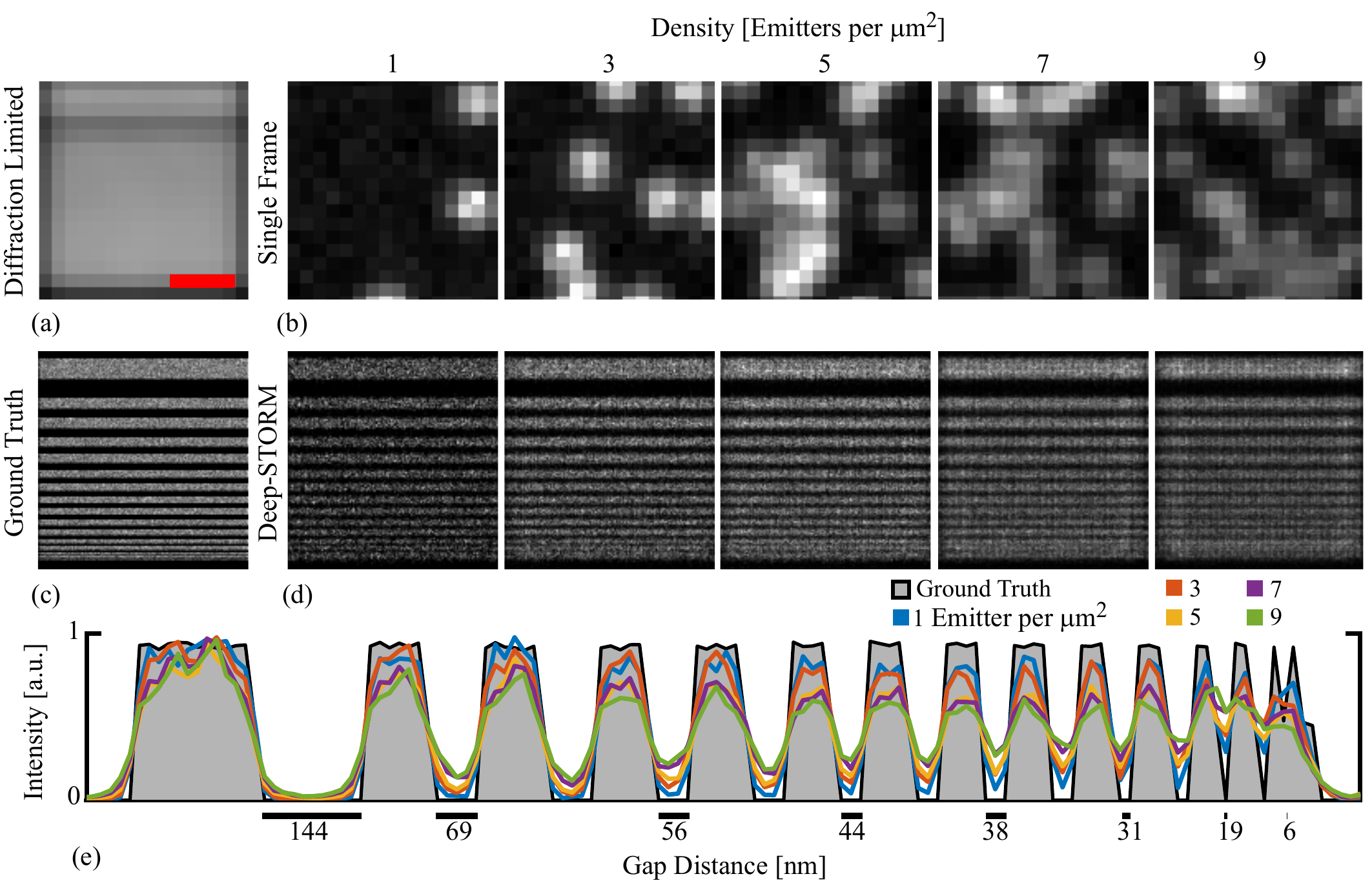}
\caption{Resolution and emitter density (simulation). (a) Diffraction-limited image of horizontal lines, scale bar 500 nm. (b) Simulated single-frames of emitters at various densities with a mean of 10 background photons per pixel and 1000 signal photons per emitter. (c) The ground truth positions of simulated emitters. (d) Deep-STORM reconstructed images. (e) Sum along the horizontal axis of the reconstruction intensities.}
\label{fig:DensityResolution}
\end{figure*}

\subsubsection{Loss Function}
Unlike typical localization-microscopy approaches, Deep-STORM directly outputs the super-resolved images rather than a localization list. Therefore, as a loss function for training the net, we adapt a regression approach. Specifically, we measure the squared $\ell_2$ distance between the network's prediction and the ground truth image (consisting of delta functions in the emitter positions) convolved with a small 2D Gaussian kernel. To promote sparsity of the network's output, we also introduce an $\ell_1$ penalizer. Let $x_i$ be the image with delta functions at the ground truth positions, $\hat{x}_i$  be the network's prediction, $g$ the Gaussian kernel, $N$ the number of images in the training set, and $\circledast$ denote convolution, then the resulting loss function is:

\begin{equation}
\ell\left(x,\hat{x}\right) = \frac{1}{N} \Sigma_{i=1}^N {\|\hat{x}_i \circledast g - x_i \circledast g\|_2^2 + \| \hat{x_i} \|_1}
\label{eq:loss}
\end{equation}

It is possible to incorporate a regularization parameter to the  $\ell_1$  term controlling the desired sparsity level; however, we observed high robustness of the resulting predictions to such a parameter. Hence, we chose to keep Deep-STORM parameter free.
The network was implemented in Keras \citep{chollet2015keras} with a TensorFlow \citep{tensorflow2015-whitepaper} backend. We trained the network for 100 epochs on batches of 16 samples using the the Adam \citep{kingma2014adam} optimizer with the default parameters, a Gaussian kernel with $\sigma=1$ pixel, and an initial learning rate of $0.001$. Training and evaluation were run on a standard work-station equipped with $32$ GB of memory, an Intel(R) Core(TM) $i7-8700$, $3.20$ GHz CPU, and a NVidia GeForce Titan Xp with $12$ GB of video memory. full network training took $\sim 2$ hours. Our code is made publicly available  \cite{Nanobioo28:online}.

\subsection{Microscopy}
Quantum dot (QD) samples were prepared by diluting 705 nm-emitting QDs (Invitrogen) $1:1000$ v/v in $1 \%$ poly(vinyl alcohol) (Mowiol 8-88, Sigma-Aldrich), then spin coating onto a standard glass coverslip (no. 1.5, Fisherscientific). Images were recorded using Nikon Imaging Software which controlled a standard inverted microscope (Eclipse TI2, Nikon) with a 405 nm light source (iChrome MLE, Toptica). Fluorescence emission from the QDs was collected using a high numerical aperture ($1.49$), $100 \times$ objective lens (CFI Apochromat TIRF 100XC Oil, Nikon), chromatically filtered to remove background (ZT488rdc $\&$ ET500LP, Chroma), then captured with a 400 ms exposure time on an sCMOS camera (95B, Photometrics). To achieve a variety of SNRs and emitter densities, images were taken with various laser powers and combined in post processing.

\section{Results}

We validated Deep-STORM on both simulated and experimental data. All microtubule reconstructions were obtained on a grid with a $12.5 \ nm$ pixel size, and QD reconstructions were obtained on a grid with a $13.75 \ nm$ pixel size. In order to estimate the expected resolution of the net's output, we simulated reconstruction of a synthetic structure of horizontal stripes at decreasing separations, on various emitter densities, using nets that were trained accordingly, for a reasonable single-molecule level SNR of $1000$ signal photons and $10$ background photons per pixel (Fig. \ref{fig:DensityResolution}). Notably, the minimal resolvable distance between stripes increases as a function of emitter density, ranging from at least ~19 nm for $1 \ [\frac{emitter}{\mu m^2}]$ to ~31 nm for $9 \ [\frac{emitter}{\mu m^2}]$. A similar resolution analysis for various SNR values is included in the Supplementary Information section.

\begin{figure*}[htbp]
\centering
\includegraphics[width=\linewidth]{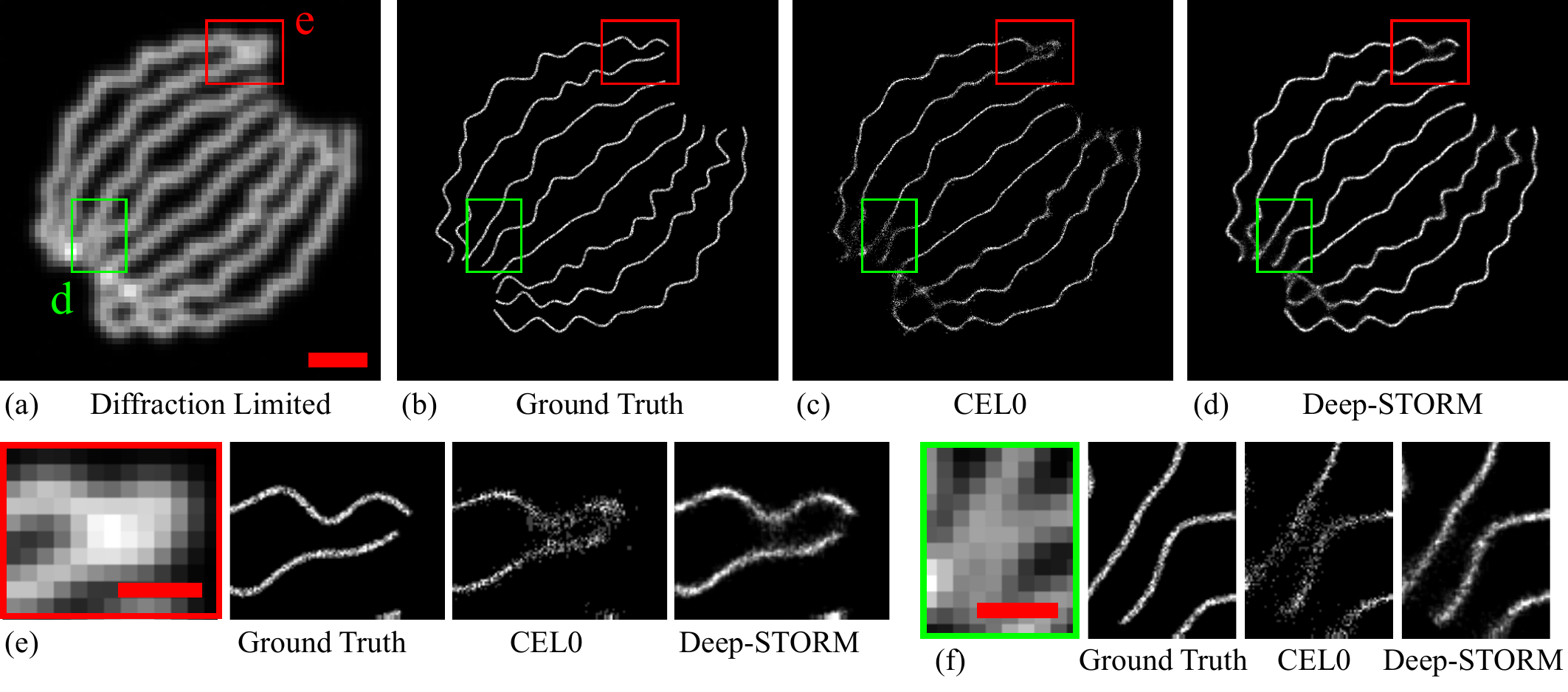}
\caption{Simulated microtubules. (a) Sum of the acquisition stack. Scale bar is $1 \ \mu m$. (b) Ground truth. (c) Reconstruction by the CEL0 method (d) Reconstruction by Deep-STORM. (e)-(f) Magnified views of two selected regions. Scale bars are $0.5 \ \mu m$.}
\label{fig:SimulatedTubulin}
\end{figure*}

\begin{figure}[htbp]
\centering
\includegraphics[width=\linewidth]{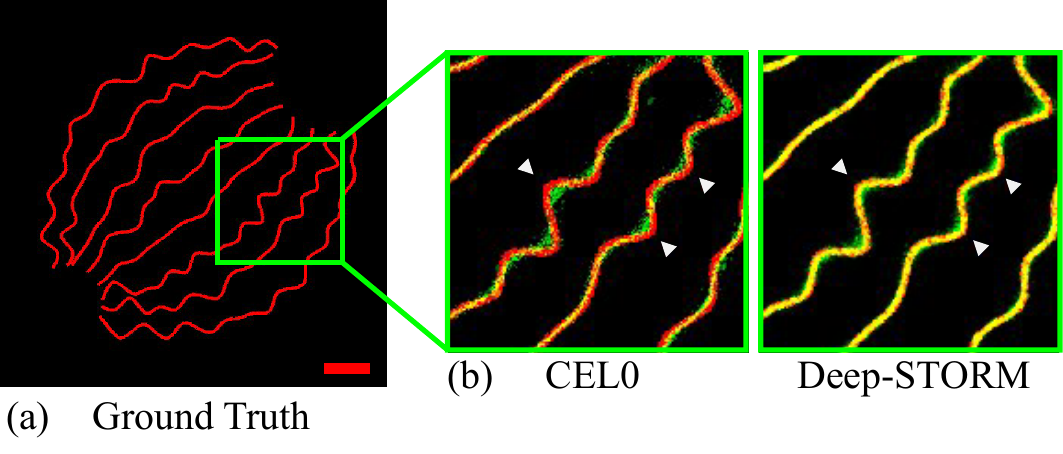}
\caption{Reconstruction accuracy. (a) Ground truth image of simulated microtubules. Scale bar is $1 \ \mu$m. (b) Merged reconstruction with the ground truth. Red shows the ground truth, green corresponds to the recovery result, and yellow marks their overlap. Note that CEL0 (left) does not follow the twisted shape in all places (white triangles), while Deep-STORM (right) better recovers the underlying structure.}
\label{fig:Curvature}
\end{figure}

Next, we tested Deep-STORM on super-resolution data, and benchmarked against a recently developed high-performance multi-emitter fitting algorithm (CEL0 \citep{Gazagnes2017}). First, we reconstructed a simulated microtubule dataset available on the EPFL SMLM challenge website \citep{Sage2015} (Fig. \ref{fig:SimulatedTubulin}). The optimal regularization parameter for CEL0 was set empirically to $\lambda=0.25$ through a comprehensive trial and error process, such that spurious detections are minimized, and the number of recovered positions was roughly equal to the number of underlying emitters. The number of IRL1 and inner FISTA iterations was set to $200$. Since Deep-STORM is not constrained to output emitter positions, we quantified the quality of the results based on image similarity measures, rather than a point-list comparison e.g. Jaccard index. Specifically, we used the standard normalized mean square error: $NMSE\left(\hat{x},x\right) = \frac{\|\hat{x} - x\|_2^2}{\|x\|_2^2}\times100\%$. Deep-STORM showed improved NMSE of $37\%$ compared to $72\%$ for CEL0 raw histogram, and $69$ for CEL0 result convolved with a Gaussian with $\sigma = 1$ pixel, where $\sigma = 1$ was optimized to produced the lowest NMSE. Deep-STORM managed to resolve nearby microtubule edges (Fig. \ref{fig:SimulatedTubulin}) and accurately recovered the underlying structure curvature compared to CEL0 (highlighted in white arrows in Fig. \ref{fig:Curvature}). To quantify the resolution, we analyzed simulated frames containing many molecules along a line, and used the trained net from above. The line width (FWHM) was $24$ nm (Supplementary figure).

\begin{figure*}[htbp]
\centering
\includegraphics[width=\linewidth]{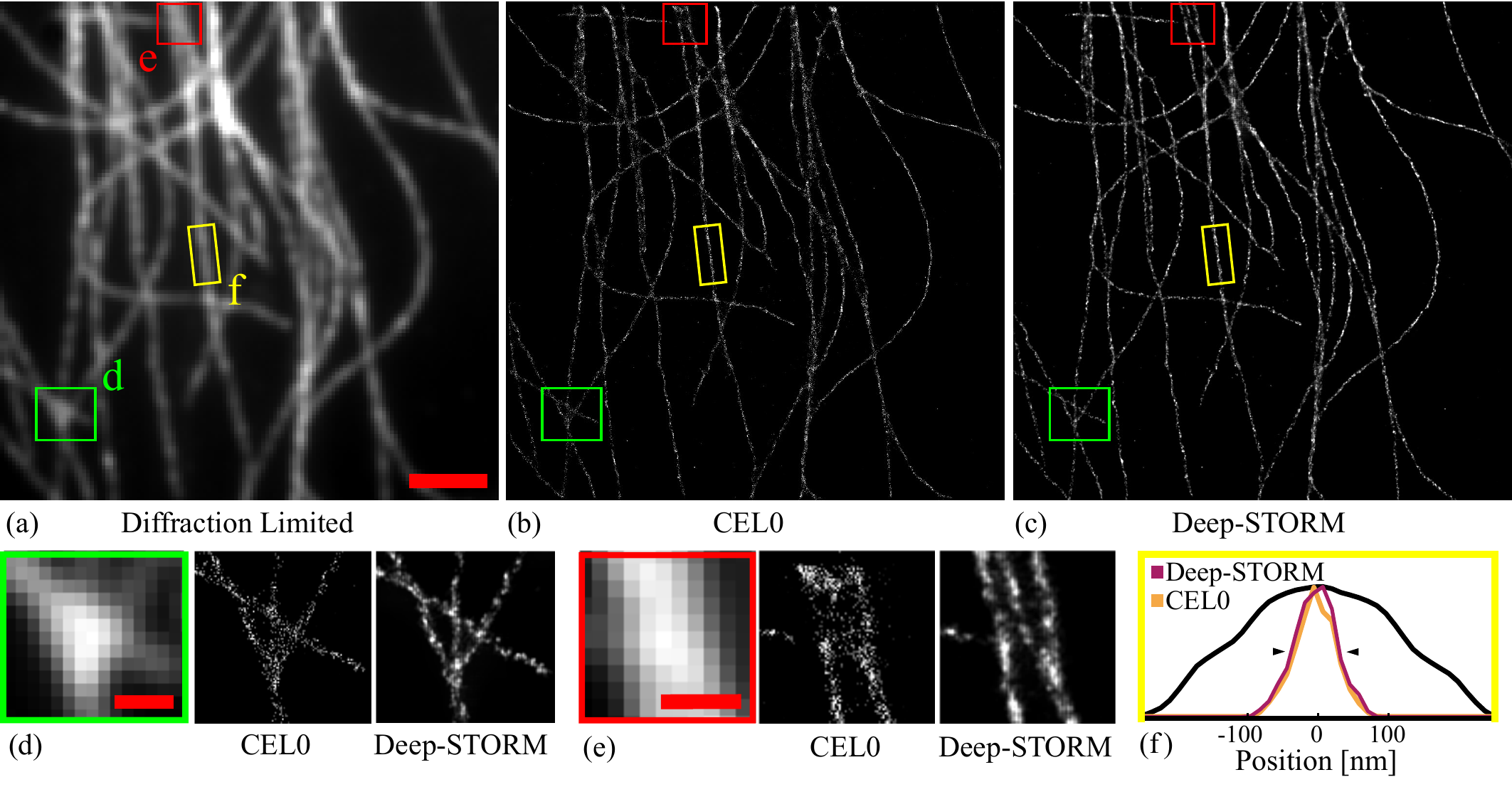}
\caption{Experimentally measured microtubules. (a) Sum of the acquisition stack. Scale bar is $2 \ \mu m$. (b) Reconstruction by the CEL0 method. (c) Reconstruction by Deep-STORM. (d)-(e) Magnified views of two selected regions. Scale bars are $0.5 \ \mu m$. (f) The width projection of the highlighted yellow region. The attained FWHM (black triangles) for CEL0 was $61 \ nm$ and  $67 \ nm$ for Deep-STORM. The black line shows the diffraction-limited projection.}
\label{fig:RealTubulin}
\end{figure*}

Second, we tested the result of Deep-STORM on experimental data obtained from Sage et al. \citep{Sage2015}, training solely on simulated data with similar experimental conditions - namely, SNR and emitter density. Deep-STORM resolves nearby lines and fine structures, and produces more continuous shapes compared to the output of CEL0 (Fig. \ref{fig:RealTubulin}). Both simulated and experimental datasets were also compared to a fast multi-emitter fitting algorithm (FALCON \citep{Min2015}). The results show that Deep-STORM is also superior to FALCON on both datasets, with $37\%$ compared to $61\%$ NMSE on the simulated data-set, and better resolved structures in the experimental data-set (Supplementary Information). 

\begin{figure}[htbp]
\centering
\includegraphics[width=\linewidth]{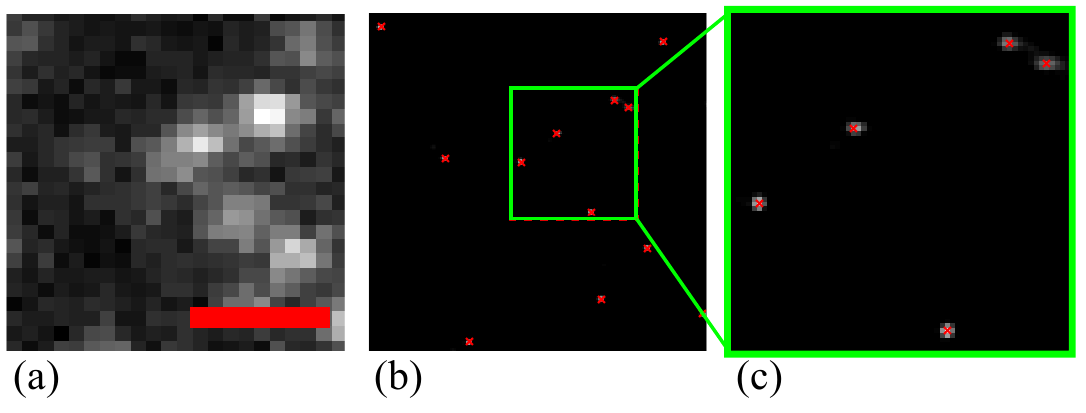}
\caption{Quantum dot experimental data. (a) Acquired low resolution image. Scale bar is $1 \ \mu m$. (b) Deep-STORM reconstruction with ground truth emitter positions (red crosses). (c) Magnified view of the a selected region in (b).}
\label{fig:QuantomDots}
\end{figure}

Ultimately, the best training set should include the aberrations in the experimental imaging system; however, very large data sets are typically used to train a deep neural network, and obtaining massive amounts of experimental images is not straight-forward. However, we found that a reasonable number of experimental images are sufficient to train a high-quality net. We trained and tested Deep-STORM on a sample containing randomly scattered fluorescent quantum dots to evaluate the performance on experimental data with a variety of SNR conditions encountered in single-molecule data sets, and at high density. To obtain a high-density data set with relatively well-known positions: we first acquired 100 images of \textit{sparse}, randomly-distributed quantum dots (a total of 1560 emitters); then localized them with high precision using ThunderSTORM \citep{Ovesny2014}. The sparse frames were next cropped into smaller regions and summed to generate dense regions for training (1200 regions) or evaluation (360 regions). Specifically, we chose 8 random regions at a time, and summed them. Notably, by combining and shifting portions of only 100 images, we produced a library of $10K$ summed regions that was used to train the network, and $3K$ for testing. The resulting imaging conditions were challenging: the emitter density of the regions was around $2 \left[ \frac{emitter}{\mu m^2}\right]$, there were ~2500 mean signal photons per emitter, and total additive gaussian noise with a standard deviation of $\sigma=20$ photons per pixel. In the $3K$ regions reserved for evaluation, Deep-STORM correctly identified $96 \%$ of emitters localized by ThunderSTORM prior to combining frames, with a low false positive rate of $1.6 \%$ (Fig. \ref{fig:QuantomDots}). 
In these conditions, Deep-STORM generates super-resolved images containing small “blobs”, usually in $3\times3$ pixel regions, with the peak being at the center. For nearby emitters, Deep-STORM produces a slightly asymmetric blob. The minor blur is also apparent in the previous examples, however it has little effect on the resulting super-resolved image (e.g. see Fig. \ref{fig:RealTubulin}).

Comparing to reconstruction of the same images using a net trained on simulated data (as described above), we found that the experimentally-trained net outperforms the simulated net, detecting $96 \%$ compared to $88 \%$ of the emitters, with a reduced false positive rate of $1.6 \%$ compared to $8.7 \%$. This test demonstrates that while simulated data can serve as excellent training data - experimentally obtained images are even better. Additionally, a high-quality reconstruction net can be trained using a small number of experimentally measured images.

Finally, we tested the robustness of our method to a mismatch between the training data and the measured image. We found that Deep-STORM is relatively robust to a density mismatch of $ \sim 2 \  \left[\frac{emitters}{\mu m^2}\right]$. In addition, we found that in case of a mismatch in SNR, it is preferable to train on lower background examples to prevent a high false positives rate (Supplementary Information).

Deep-STORM not only yields image reconstruction results that are comparable to or better than leading algorithms, but also does so $\sim 1-3$ orders of magnitude faster. Table \ref{tab:running-time} compares the run time of Deep-STORM vs. CEL0 and FALCON on both simulated and experimental  microtubule datasets (Figs. \ref{fig:SimulatedTubulin} $\&$ \ref{fig:RealTubulin}). The simulated dataset consists of 361 frames containing $\sim 81K$ emitters in total, with mean density of $5.48 \ [\frac{emitter}{\mu m^2}]$. The experimental dataset consists of 500 frames containing $\sim 520K$ emitters, with mean density of $6.35 \ [\frac{emitter}{\mu m^2}]$, approximated using the number of emitters recovered by CEL0. Deep-STORM exhibits significantly superior runtime, especially when introducing GPU acceleration, equivalent to localizing $\sim 20000$ emitters per second, compared to $\sim 1500$ emitters per second by the fastest existing multi-emitter fitting method to our knowledge (FALCON \citep{Min2015}). 

\begin{table}[!htb]
\centering
\caption{\bf Runtime comparison}
\resizebox{\columnwidth}{!}{%
\begin{tabular}{|c|c|c|c|c|}
\hline
Dataset & $Grid \ size$ & CEL0 & FALCON & Deep-STORM \\
& & CPU [s] & CPU / GPU [s] & CPU / GPU [s]\\
\hline
Sim. & $512\times 512$ & $18677$ &  $1465 / 122$ & $123 / 4$\\
\hline
Exp. & $1024\times 1024$ & $175200$ & $10177 / 434$ & $715 / 27$\\
\hline
\end{tabular}}
  \label{tab:running-time}
\end{table}

\section{Discussion}
Since the introduction of single-molecule localization microscopy, numerous algorithms have been developed to reconstruct super-resolved images from movies of stochastically-blinking emitters. 

In particular, considerable effort has been invested in solving the high-density emitter-fitting problem. Indeed, current methods for multi-emitter fitting produce high-quality images; however, this comes at a high computational cost, i.e. runtime, as well as frequently necessitating parameter-tuning. In this work, we have presented a fast, precise, and parameter-free method for super-resolution imaging from localization-microscopy type data. Deep-STORM uses a convolutional neural network trained on easily-simulated or experimental data.

Our experiments show that the net used in this work performs well up to a density of $\sim 6 \ [\frac{emitter}{\mu m^2}]$, which is similar to leading multi-emitter fitting methods, after tuning their parameters accordingly. We note that, in general, the maximal allowable density would depend also on SNR. Notably, the main reason Deep Learning is highly suitable for the application presented in this work is the simplicity in which training data can be generated. Namely, single-molecule images with realistic noise models are straight forward to simulate in large numbers, which are often required in Deep Learning.

Our simulations show that Deep-STORM exhibits high robustness to emitter density and SNR used for training (Supplementary Information); nevertheless, in order to further increase performance for cases such as time-varying emitter densities or signal/background levels, the following simple generalization can be considered: Since training of the net is performed offline, a pre-training of a set of nets for various SNR and density values can be performed once. Then, in the reconstruction stage, a fast optimal selection step per-frame can be applied to each captured frame, routing it to the best net, considering the estimated SNR and emitter density of the current frame.

Although Deep-STORM uses localization-microscopy type movies to produce a super-resolved image, it is not a localization based technique. Localization microscopy is based on the additional information inherent in blinking molecules. However, as was demonstrated by other techniques, e.g. SOFI \citep{Dertinger2009}, extracting this information does not necessarily require compiling a list of molecular positions. Deep-STORM implicitly uses this additional information content to directly reconstruct a super-resolved image. The technique combines state-of-the-art resolution enhancement, unprecedented speed, and high flexibility (parameter-free operation). This combination produces a technique capable of video-rate analysis of super-resolution localization-microscopy data that requires no expertise from the end user, overcoming some of the the most significant limitations of existing localization methods.

\bibliographystyle{ieeetr}
\bibliography{main_new}

\section*{Funding Information}
E.N. is supported by a Google research award. L.E.W. and Y.S. are supported by the Zuckerman Foundation, Y.S. is supported in part by a Career Advancement Chairship from the Technion, Israel Institute of Technology. T.M. is supported in part by the Ollendorff Foundation, the Taub Foundation (through a Horev fellowship), an Alon Fellowship, and the Israel Science Foundation (grant No. 852/17). We gratefully acknowledge the support of NVIDIA Corporation with the donation of the Titan Xp GPU used for this research. 

\section*{Acknowledgements}
The authors thank Dr. Daniel Freedman of Google Research for fruitful discussions.

\section*{Supplemental Documents}
See supplementary material for supporting information.

\end{document}